\documentclass[twocolumn]{aastex631}

\begin{document}

\title{Jupiter Atmospheric Models and Outer Boundary Conditions for Giant Planet Evolutionary Calculations}

\shorttitle{Atmospheric Boundary Conditions of Jupiter with Irradiation and Clouds}
\shortauthors{Chen et al.}


\author[0000-0003-3792-2888]{Yi-Xian Chen}
\affiliation{Department of Astrophysical Sciences, Princeton University,  4 Ivy Lane, Princeton, NJ 08544, USA}

\author[0000-0002-3099-5024]{Adam Burrows}
\affiliation{Department of Astrophysical Sciences, Princeton University,  4 Ivy Lane, Princeton, NJ 08544, USA}

\author[0000-0001-6635-5080]{Ankan Sur}
\affiliation{Department of Astrophysical Sciences, Princeton University,  4 Ivy Lane, Princeton, NJ 08544, USA}

\author[0000-0001-6708-3427]{Roberto Tejada Arevalo}
\affiliation{Department of Astrophysical Sciences, Princeton University,  4 Ivy Lane, Princeton, NJ 08544, USA}




\begin{abstract}
We present updated atmospheric tables suitable for calculating the post-formation evolution 
and cooling of Jupiter and Jupiter-like exoplanets. 
These tables are generated using a 1D radiative transfer modeling code that incorporates the latest opacities and realistic prescriptions for stellar irradiation and ammonia clouds. 
To ensure the accuracy of our model parameters, 
we calibrate them against the measured temperature structure and geometric albedo spectrum of Jupiter, 
its effective temperature, and its inferred internal temperature. 
As a test case, 
we calculate the cooling history of Jupiter using an adiabatic and homogeneous interior and compare with extant models now used to evolve Jupiter and the giant planets. 
We find that our model reasonably matches Jupiter after evolving a hot-start initial condition to the present age of the solar system, 
with a discrepancy in brightness temperature/radius within two per cent. Our algorithm allows us to customize for different cloud, irradiation, and metallicity parameters. 
This class of boundary conditions can be used to study the evolution of solar-system giant planets and exoplanets with more complicated interior structures and non-adiabatic, inhomogeneous internal profiles.
\end{abstract}



\section{Introduction} \label{sec:intro}

The cooling and contraction of the 
mostly convective interiors of giant planets occur after their formation by core accretion \citep{Pollack1996,DAngelo2003,Li2021} or disk instability \citep{Boss1997}, and is regulated by the radiative properties of the thin layers of atmosphere on the planets' surface. 
To determine the external boundary conditions for time-dependent evolutionary calculations, 
it is crucial to generate a set of model atmospheres at various values of surface gravity and internal emission and/or entropy. 
The interior models, 
which provide information on the thermodynamic properties and composition of the planet, 
can therefore be connected over time using boundary conditions obtained through interpolation from this grid of models.

Pioneering investigations into the evolution of Jupiter and Saturn, such as \citet{Graboske1975}, \citet{Pollack1977}, 
and \citet{Hubbard1977}, were conducted before the discovery of hot Jupiters in 1995 \citep{Mayor1995}. Subsequently, 
new sets of models were developed to cover a wider range of parameters motivated by the need to study exoplanets \citep{Burrows1997,Burrows2001}, and include better prescriptions for irradiation \citep{Sudarsky2000,Burrows2006}, 
opacities \citep{SharpBurrows2007}, 
and cloud formation \citep{Burrows2006cloud}. 
These improvements have resulted in more refined atmospheric boundary conditions for planetary evolution, 
including ones customized for specific extrasolar sources \citep[e.g.][]{Burrows2003} 
or solar system giant planets \citep{Fortney2011}.

Emission spectra of gas giants produced in self-consistent atmosphere calculations can be directly compared with observed spectra of giant planets to constrain their properties \citep{Sudarsky2003}. 
The launch of JWST has ushered in a new era of exoplanet atmospheric detection with unprecedented precision \citep{JWST1,JWST2}, 
requiring more complete archives of reference theoretical spectra. In this sense, modeling the outer atmospheres of gas giants and generating tabulated spectra are critical for two main reasons.  First, theoretical models are necessary to interpret important physical quantities, 
such as temperature, gravity, radius, composition, and albedo, 
from spectra obtained through direct detection. 
Second, theoretical spectral models inform observers concerning which regions of the spectrum are most likely to yield the most insight into a planet's atmosphere. 
Moreover, spectrum calculations provide more well-calibrated atmospheric boundary conditions, which can be combined with updated knowledge of opacity and interior equations for evolutionary models that constrain the age of directly imaged planets.

This paper is structured as follows: In \S\ref{sec:model_atmos}, we describe the numerical methods used to calculate our models, focusing on additions and alterations, such as ammonia clouds and irradiation, and compare to previous models for gas giant evolution \citep{Burrows1997,Fortney2011}. 
In \S \ref{sec:jupiter}, we calibrate a fiducial set of numerical parameters, including irradiation redistribution and cloud modal size, based not only on Jupiter's metallicity ($\sim$3$\times$ solar), gravity, and measured effective temperature, but also on observed albedo and temperature-pressure profiles. 
In \S \ref{sec:evo_variables}, we calculate the cooling of gas giants using a standard interior equation of state, and illustrate the effects of cloud and irradiation parameters on the evolution of gas-giant radii, the internal temperature ($T_{\rm int}$), and the effective temperature ($T_{\rm eff}$). 
Finally, in \S \ref{sec:discussions}, we discuss our conclusions and their implications for the next generation of giant-planet evolutionary models that incorporate more sophisticated thermal and compositional profiles and energy transport modalities.

\section{Model Atmospheres}
\label{sec:model_atmos}

We generate model atmospheres using the 1D atmosphere and spectral code \texttt{CoolTLusty} \citep{Hubeny1995,Sudarsky2003,Sudarsky2005,Burrows2008}. 
This code self-consistently calculates the spectrum and structure of a plane-parallel atmosphere with radiative transfer, 
given a detailed suite of opacity data compiled into thermochemical equilibrium tables \citep{BurrowsSharp1999, SharpBurrows2007}. The opacities are treated using line-by-line sampling and have recently been significantly updated in \citet{Lacy2023} (see their Appendix A). 
Their strategy for calculating absorption cross sections is mainly guided by recent progress by the ExomolOP \citep{Chubb2021} and EXOPLINES \citep{Gharib-Nezhad2021} collaborations. 
A publicly available Fortran code, 
{\em exocross} \footnote{https://exocross.readthedocs.io/en/latest/}, was employed to compute absorption cross sections over a grid of temperatures, pressures, and wavenumbers.  
User-defined aspects of {\em exocross} absorption cross-section calculations include line lists, line profiles  
with pressure-dependent broadening if desired (the Voigt profile is usually applied), line-wing cutoffs, and an optional line strength threshold. For each molecule, 
the line lists recommended by the ExoMol team were adopted,
with additional updates to cover higher temperature and larger wavelength ranges.  The most relevant updated molecules
include water, ammonia, methane, and molecular hydrogen. 
The line list for molecular ${\rm CH}_4$ \citep{Yurchenko2014} is additionally supplemented at shorter wavelengths by cross sections inferred from Jupiter’s spectrum in \citet{Karkoschka1994}.

In our atmospheric calculations, we use 100 atmospheric layers, 
and 5000 frequency points spaced evenly in log from 0.5 to 300 $\mu$m. 
We neglect disequilibrium chemistry discussed in \citet{Lacy2023}, 
but add treatments for stellar irradiation (\S\ref{irrad}). 
For fiducial evolutionary calculations of a gas giant in isolation, the metallicity ($Z$), the intrinsic temperature ($T_{\rm int}$; or equivalently the internal flux) and outer planet radius are needed to construct the thermal boundary condition and calculate the specific entropy at the base of the atmosphere's radiative zone. 
Inverting this table, 
$T_{\rm int}$ becomes a function of the surface gravity and entropy per baryon ($S$) in the deep interior, sometimes parameterized by $T_{10}$, the temperature that the interior isentrope would have if extrapolated to a pressure of 10 bars \citep{Burrows1997,Hubbard1999,Fortney2011}. 
When there is no external stellar irradiation source, 
the total effective temperature ($T_{\rm eff}$) is equal to $T_{\rm int}$, which is why in studies of isolated gas giant evolution $T_{\rm eff}$ is often simply used in the place of $T_{\rm int}$ \citep[e.g.][]{Burrows1997}.

When stellar irradiation is taken into account, 
$T_{\rm eff}$ will differ from $T_{\rm int}$ due to the fraction of absorbed stellar irradiation by the planet. 
Though we do not use such a quantity, 
this fraction is often, though crudely, identified with a $T_{\rm eq}$, such that $T^4_{\rm eff}$ is set equal to $T^4_{\rm eq}$ + $T^4_{\rm int}$ \citep{Saumon1996,Sudarsky2000}. In \texttt{CoolTLusty}, 
the wavelength-dependent geometric albedos are calculated self-consistently and the absorbed stellar heat is intrinsically accounted for \citep{Sudarsky2003,Burrows2006}. 
Although only $T_{\rm int}$ is relevant to the internal evolution,  
$T_{\rm eff}$ can be obtained from the total emission and compared to measured values for Jupiter. 
Hence, when we calculate evolutionary tracks for $T_{\rm int}$, 
we simultaneously obtain the associated $T_{\rm eff}$ evolution, along with the frequency-dependent spectra.

\subsection{Cloud Models}
\label{cloud}

Since we focus on gas giants with effective temperatures of $80$ to $200$ K, 
ammonia clouds will form in the atmosphere and have an impact on their late-time evolution. 
Such clouds are a much-studied feature of both Jupiter and Saturn \citep{Brooke1998,Sromovsky2018,dePater2019}.
Water clouds appear earlier when the atmospheric temperatures go below $\sim$400 K, but are quickly buried deep early on. 
Young, more massive, giant planets should evince water clouds that for them would need to be included in evolutionary calculations. 
However, such is not the case for the current Jupiter and Saturn and we ignore them here. 
Our treatment of cloud opacity is the same as in \citet{Lacy2023}, and we use the same cloud shape parameters 
(the compact ``E"-type) and supersaturation factor (0.01). 
We adopt the Clausius-Clapeyron line for the base of the cloud. 
The cloud spatial distributions we employ are elaborated in earlier works that published cloudy atmospheric models using \texttt{CoolTLUSTY} \citep{Burrows2006cloud,Madhusudhan2011}.
We assume the cloud species to have a Deirmendjian size distribution \citep{Deirmendjian1964}:

\begin{equation}
    n(a) \propto \left(\frac{a}{a_0}\right)^6 \exp\left[-6\left(\frac{a}{a_0}\right)\right]\, ,
\end{equation}

with a default modal size ($a_0$) on the order of microns.
A Deirmendjian particle size distribution reproduces that of the Earth's clouds for a $a_0$ of 4$\mu$m. As we show below (\S \ref{sec:jupiter}), 
$a_0 \sim 1\mu$m is consistent with the Bond and geometric albedos of Jupiter. 
However, we also experiment with larger particle sizes to explore the model dependencies. 
For a given particle size distribution, 
the frequency-dependent absorption cross section $\sigma(\nu, a_0)$ is then calculated using Mie theory \citep{Kerker1969} and converted into an opacity per unit mass \citep{Lacy2023}. 

\subsection{Irradiation}
\label{irrad}

We treat irradiation by a sun-like star using the approach found in \citet{Burrows2006} and \citet{Burrows2008}. 
The solar power contribution at 5.2AU is calculated by intercepting the incident solar flux with an area of $\pi R_J^2$, 
multiplying by $f = 1/2$ to account for an average zenith angle of $60^{\circ}$ \citep{Appleby1986,Marley1999,Fortney2011}. 
Note that though \citet[][see their Appendix D]{Burrows2008} 
proved that one should apply a factor of 2/3 in the limit of strongly irradiated Hot Jupiters, 
a factor of 
1/2 may be more appropriate in the case of gas giants with moderate to large orbital distances. 
{Therefore, the frequency-integrated stellar flux expressed via the $H$-moment is}
\begin{equation}
    H_{\rm ext} = f \left(\frac{R_*}{d}\right)^2 \frac{\sigma}{4\pi} T_*^4,
\end{equation}
{where $R_*$ is stellar radius, $d$ is the star-planet distance, and $T_*$ is the effective temperature of the stellar surface.}

{Furthermore, 
we apply a redistribution parameter $P_{\rm irr}$ to represent the fraction of the flux that is redistributed. 
Effectively, 
our treatment removes a fraction of the irradiation $H_{\rm irr} = P_{\rm irr} H_{\rm ext}$ from the day side (and constitutes an energy input to the night side). 
Although all profiles we calculate in this paper belong to the day side, 
it's worth noting that conventional setups with no redistribution ($P_{\rm irr}=0$) imply zero redistributed heating to the night side. }

{To redistribute heat from the day side, we add the additional sink term $-D$ to the radiative transfer equation, expressed in terms of the integrated column mass of the atmosphere $m$:}

\begin{equation}
    D(m) = \frac{2H_{\rm irr}}{m_1-m_0} \frac{m_1-m}{m_1-m_0}
\end{equation}
{where the parameters $m_1$ and $m_0$ are column masses corresponding to the limiting pressures $P_0$ and $P_1$ (the sink term is set to zero outside this mass/pressure range), 
such that $D(m)$ linearly decreases with $m$, achieving a value of zero at the bottom of the redistribution zone ($P_1$). For more details, see Appendices A \& B of \citet{Burrows2008}. Integrating $D(m)$ over the column mass confined between the two limiting pressures gives $H_{\rm irr}$. We restrict the redistribution altitudes to be between $P_0 = 0.05$ and $P_1 = 0.5$ bar as limiting pressures, guided by the temperature structure of Jupiter. }

As we show below,  larger $P_{\rm irr}$ leads to a steeper temperature inversion in the stratosphere, 
and we select a fiducial $P_{\rm irr}$ of $0.15$ based on calibration with measured Jupiter temperature-pressure profiles. 


\section{Calibration with Jupiter}
\label{sec:jupiter}

\subsection{Fitting the Atmospheric Temperature and Geometric Albedo Profiles}
\label{fitting}

Here, we determine a set of fiducial parameters for the incorporation of the effects of clouds and irradiation into our atmosphere boundary conditions by constraining our models to fit observation data, 
specifically the measured albedo and temperature-pressure (TP) profiles of Jupiter \citep{Seiff,Karkoschka1998}. 
For this purpose, 
we fix $\log(g[{\rm cgs}])=3.4$ and the metal abundance ($Z$) at 3.16 times solar \citep{Niemann1998}. 
There are some uncertainties in the helium abundance measurements and the fraction may also vary with time in more modern evolutionary calculations, 
but we choose $Y=0.25$ as the fiducial value and have tested that all outcomes are quite insensitive to \textit{atmospheric} $Y$ values between 0.22 and 0.28.
In models with irradiation, 
we assume the central stellar source of irradiation has a blackbody spectrum with a temperature of $T_*=5777$K, radius of $R_\odot$, and is a distance of $d = 5.2$AU from the planet. Furthermore, as has been traditional in the literature, we employ the {``pseudo" effective temperature} $\tilde{T}_{\rm eff}$ as a constraint to help determine consistent values of $T_{\rm int}$ for each set of cloud or irradiation parameters. 
More specifically, we use as a fiducial $\tilde{T}_{\rm eff} = 125.57\pm 0.07$K, based on Cassini CIRS and VIMS observations \citep{Li2012}. These authors found that while infrared emissions measured using the CIRS (Composite Infrared Spectrometer, $> 7\mu$m) account for most of Jupiter's emissions, the emissions around 5$\mu$m measured in the VIMS band (Visual and Infrared Mapping Spectrometer, 4.5-5.5 $\mu$m) have a non-negligible $\sim 1\%$ contribution to the total emission flux, whose specific fraction varies with latitude. 
To be consistent in matching this constraint, 
our $\tilde{T}_{\rm eff}$ is defined as the sum of the thermal emission larger than 4.5 microns, 
to differentiate with ${T}_{\rm eff}$, 
the total integrated brightness temperature in theoretical models. {Generally, $\tilde{T}_{\rm eff}$ varies within a factor of 2\% from ${T}_{\rm eff}$ and does not constitute a major difference from previous evolution models of irradiated Jupiters. Moreover, as we will elaborate below, only the tabulated values of convective zone entropy $S$ will determine the evolution track, and not $\tilde{T}_{\rm eff}$}. For the purpose of generating an atmospheric boundary condition calibrated on measurements of Jupiter, 
we explore the parameter space of $P_{\rm irr}$ and $a_0$ of models with $\tilde{T}_{\rm eff}\approx 125.57$K, 
and converge (see below) on a best-fit model with $P_{\rm irr} = 0.15$ and $a_0 = 1\mu$m. 

\begin{figure}[htbp!]
\centering
\includegraphics[width=0.45\textwidth,clip=true]{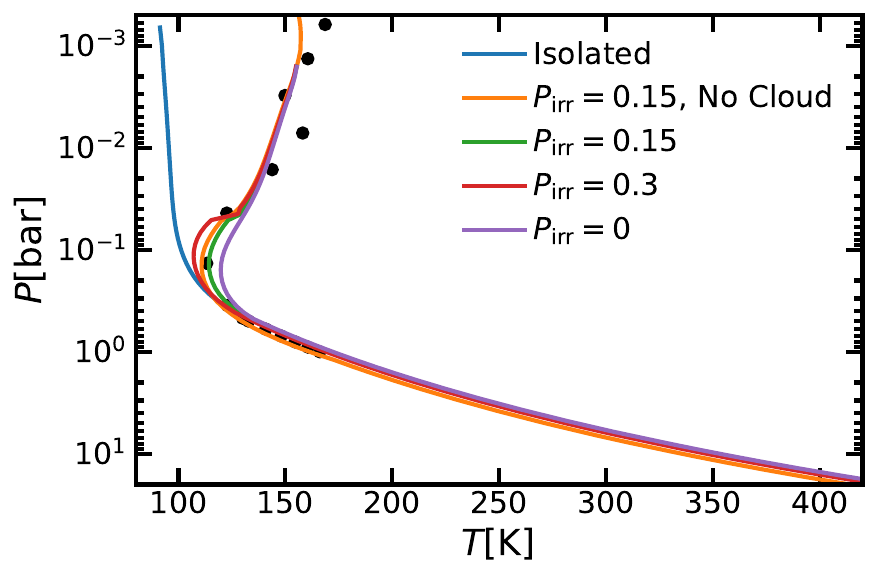}
\caption{Comparison of the temperature-pressure profiles of different Jupiter models at $\tilde{T}_{\rm eff} = 125$K. Only the isolated (no irradiation or cloud) models show no temperature inversion. 
All other models are irradiated at 5.2 A.U., either without clouds or with clouds of modal particle size of $1\mu$m. We observe that $P_{\rm irr}$ is closely related to the temperature valley at the inversion around 0.1 bar, 
and that $P_{\rm irr}=0.15$ best reproduces the observational data. 
Once the redistribution factor is set, the existence of clouds does not alter the temperature-pressure profile. Other particle sizes also have TP profiles that nearly coincide with the green line. {The Galileo entry probe data from \citet{Seiff} is plotted in black.}}
 \label{fig:PT_comparison}
\end{figure}

During this process, 
the Temperature-Pressure(TP) profile measured by the {\em Galileo} entry probe \citep{Seiff}, 
which boasts a characteristic temperature inversion, 
places a tight constraint on $P_{\rm irr}$. {This temperature inversion in the ``stratosphere" of Jupiter is also seen in Voyager data \citep{Lindal1981}, 
and attributed to the interaction between alkanes and stellar irradiation \citep{Yelle2001}. 
The lower stratosphere, where the temperature profile is relatively smooth, is mainly heated by absorption of methane of sunlight in the near-IR wavelengths. 
In our modeling, the redistribution introduces extra effective cooling between $0.05$ and $0.5$ bars.}
In Figure \ref{fig:PT_comparison}, we plot the TP profiles of characteristic atmospheric models against the {\em Galileo} entry probe data. For comparison, 
the blue line corresponds to a model with neither a cloud nor irradiation. In this model, 
$T_{\rm int} = T_{\rm eff} \approx \tilde{T}_{\rm eff}$ and no temperature inversion is observed. 
The orange line shows that the irradiation-only model with $P_{\rm irr} = 0.15$ fits reasonably well with the entry probe measurements, 
and we found that $T_{\rm int}$ needed to be $\sim$97 K to satisfy the $\tilde{T}_{\rm eff}=125$ K constraint. 
This is slightly different from the 99 K estimation of \citet{Fortney2011}, 
which may be a consequence of their neglect of redistribution. 
In Figure \ref{fig:PT_comparison}, we include three additional models with ammonia clouds of modal particle size $a_0 = 1\mu$m and different values of $P_{\rm irr}$. Generally, 
at fixed $\tilde{T}_{\rm eff}$, 
cloud existence or particle size does not have a strong impact on the TP profile since the stratosphere is not heated by clouds, 
as can be seen by comparing $P_{\rm irr}=0.15$ models with and without ammonia clouds. 
However, {the temperature profile transition at the ``tropopause"} is quite sensitive to $P_{\rm irr}$. 
While for $P_{\rm irr} = 0.3$ the thermal inversion at $\sim 0.5$ bar is too steep, 
it becomes too smooth without any redistribution cooling, 
and for $P_{\rm irr}=0.15$ the TP profile fits well for all cloud particle sizes. 
Moreover, 
a value for $T(1 {\rm bar})$ of $166\pm 1$K, consistent with Galileo entry probe data \citep{Seiff}, 
is well reproduced and comports with other previous theoretical models of Jupiter's atmospheric thermal profile \citep{Guillot2004}. {Note that our models still neglect other potentially important heating/cooling sources 
such as aerosols \cite{Zhang2015}, wave-breaking \citep{Young2005}, etc., but we emphasize that we mainly aim to improve upon past published literature interested in boundary conditions for evolutionary models.}

To break the degeneracy in the selection of cloud modal particle size, 
the geometric albedo spectrum serves as an additional useful constraint. 
Inspired by the conclusion that 1-10 micron ${\rm NH}_3$ ice particles fit the Infrared Space Observatory data \citep{Brooke1998}, 
we experimented with clouds with modal particle size ($a_0$) of 1, 3, and 10 microns. 
We note there are more recent studies, based on New Horizons LEISA (Linear Etalon Imaging Spectral Array) data, arguing that the characteristic particle size is more likely between $\sim$2 and 4 microns \citep{Sromovsky2018}, 
with a relatively wide distribution, 
and also that ${\rm NH}_4{\rm SH}$ solids can contribute to the clouds. 

\begin{figure}[htbp!]
\centering
\includegraphics[width=0.45\textwidth,clip=true]{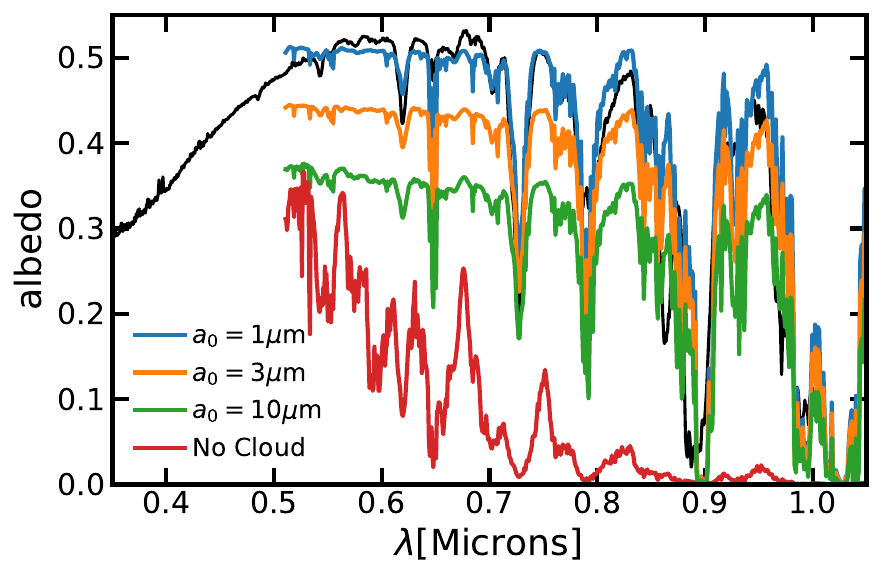}
\caption{Geometric albedo spectra for different irradiated atmospheric models without clouds (red) or with clouds of different modal particle sizes, fixing $P_{\rm irr} = 0.15$. The black spectrum is from \citet{Karkoschka1998}. 
}
 \label{fig:albedo_comparison}
\end{figure}

Exploring this relevant range of parameters, 
we compare our geometric albedo spectrum with the high resolution sub-micron geometric albedo spectrum from the European Southern Observatory \citep{Karkoschka1998}. 
Our results for different models and particle sizes are shown in Figure \ref{fig:albedo_comparison}, plotted against the data of \citet{Karkoschka1998}. We note that for the 1-micron case 
our geometric albedo is consistent with observations for $\gtrsim 0.5 \mu m$. However, the irradiation-only model (red solid line) seriously underestimates the geometric albedo profile; even the overall shape cannot be matched. 
Note that we do not attempt to model the extra scatterer/chromophores \citep{Carlson2016} below $0.5\mu$m wavelength, about which there are significant uncertainties \citep{Lacy2019}. For cloud sizes of 3 and 10 microns, although the change in particle size does not affect the temperature profile, the geometric albedo spectrum itself is slightly below the observations.

In conclusion, 
we find that the general geometric albedo spectrum can be reproduced only with the inclusion of clouds and that its values sensitively increase with decreasing cloud particle size. 
We also find that for $a_0 = 1\mu$m the albedo spectrum matches well the measured values in \citet{Karkoschka1998}. For this set of parameters, the internal temperature ($T_{\rm int}$) at a $\tilde{T}_{\rm eff}$ of $\sim125.57$ K is constrained to be $\approx$102 K, 
in contrast with a $T_{\rm int}$ of $\approx 97$ K for the no-cloud case.

\subsection{A Discussion Concerning the Bond Albedo}
\label{sec:albedo}

In theory, one can also relate the scattered/reflected fraction of solar irradiation with an effective average Bond albedo $A_B$, 
where the fraction that contributes to $\sigma T_{\rm eff}^4$ is $1-A_B$. 
\citet{Lietal2018} analyzed the Cassini VIMS and ISS (Imaging Science Subsystem) data to estimate a Bond albedo of 0.503$\pm$0.012, which is significantly larger than the previous adopted values around 0.343 \citep{Conrath1989,Fortney2003}. 
By applying the incident solar flux at Jupiter's orbital radius, they derive an internal flux of 7.485 $\pm$ 0.163 W/m$^2$, 
consistent with Jupiter's total emitted power of 14.098$\pm$0.031 W/m$^2$ (from the earlier measurement of $\tilde{T}_{\rm eff} = 125.57\pm 0.07$K in \citet{Li2012}) and corresponding to a $T_{\rm int}$ of $\approx 107$ K. As we noted, 
our cloudless model with or without redistribution, as well as the cloudless Jupiter model of \citet{Fortney2011}, 
constrained by $\tilde{T}_{\rm eff}\approx 125$ K, both give $T_{\rm int} \lesssim 100$K, which is inconsistent with the latest measurements of Jupiter's heat balance. As a matter of fact, 
this discrepancy implies there should be significant cloud effects at work scattering away a larger fraction of incident irradiation. 

However, we caution that given our planar calculations 
we have no associated phase integrals to directly provide the Bond albedo from the geometric albedo spectrum. 
The spectrum-integrated Bond albedo can be calculated from a spectroscopic and atmospheric model by integrating the monochromatic spherical albedo, weighted by the stellar flux $F_*(\lambda)$:

\begin{equation}
    A_B = \frac{\int F_*(\lambda) A_s(\lambda) d\lambda}{\int F_*(\lambda)}\, ,
\end{equation}

where $A_s(\lambda)$ is the spherical albedo (monochromatic Bond albedo). The spherical albedo and monochromatic Bond albedo are the same and given by the product of $A_G(\lambda)$ and the phase integral ($q(\lambda)$). Due to the stellar flux weighting, the optical wavelengths dominate the total Bond albedo. 
However, $q(\lambda)$ cannot be self-consistently determined from a 1D planar model, and the scattering of light over all phase angles must be studied in 2D \citep{Marley1999,Cahoy2010,Madhusudhan2012}. 

\citet{Lietal2018} measured the realistic $q(\lambda)$ for Jupiter, averaged over all phase angles, to be $\approx 1.3$ in the optical wavelength range of interest (see their Figure 3). 
This means that in realistic multidimensional models, 
we should expect a smaller fraction of the irradiation to contribute to the planet emissions calculated in 1D; 
hence, it's natural that the internal $T_{\rm int}$ is larger in their measurements than in our fiducial model (which boasts  a consistent geometric albedo spectrum but lacks the phase integral). 
We note that, informed by their model, if we multiply our $A_G$ by their factor of 1.3 to mimic a spherical albedo, we obtain a consistent frequency-integrated value for the Bond albedo of 0.5. 
However, since the phase integral is measured only for Jupiter and subject to many uncertainties, 
we make no attempt to reconcile this mismatch with artificial treatments that could lead to significant confusion 
\footnote{However, see the next section for tests varying $f$ that might inform the general dependence of $T_{\rm int}$ on the phase integral.}. The scattering phase function can be treated realistically only with multi-dimensional radiative transfer simulations. 
These uncertainties and caveats not withstanding, we conclude that our choice of fiducial parameters: 
$P_{\rm irr} = 0.15$ and $a_0 = 1\mu$m (with both irradiation and ammonia clouds included) suitably reproduces the observation data and is sufficient to provide practical atmosphere boundary conditions for the evolution of Jupiter-like planets. 
Since redistribution and clouds seem necessary to reproduce temperature inversions and the geometric albedo spectrum of Jupiter, 
the inclusion of these effects is an improvement over earlier realizations of atmospheric boundary conditions for evolutionary calculations \citep{Burrows1997,Fortney2011}. 





\begin{figure*}[htbp!]
\centering
\includegraphics[width=0.8\textwidth,clip=true]{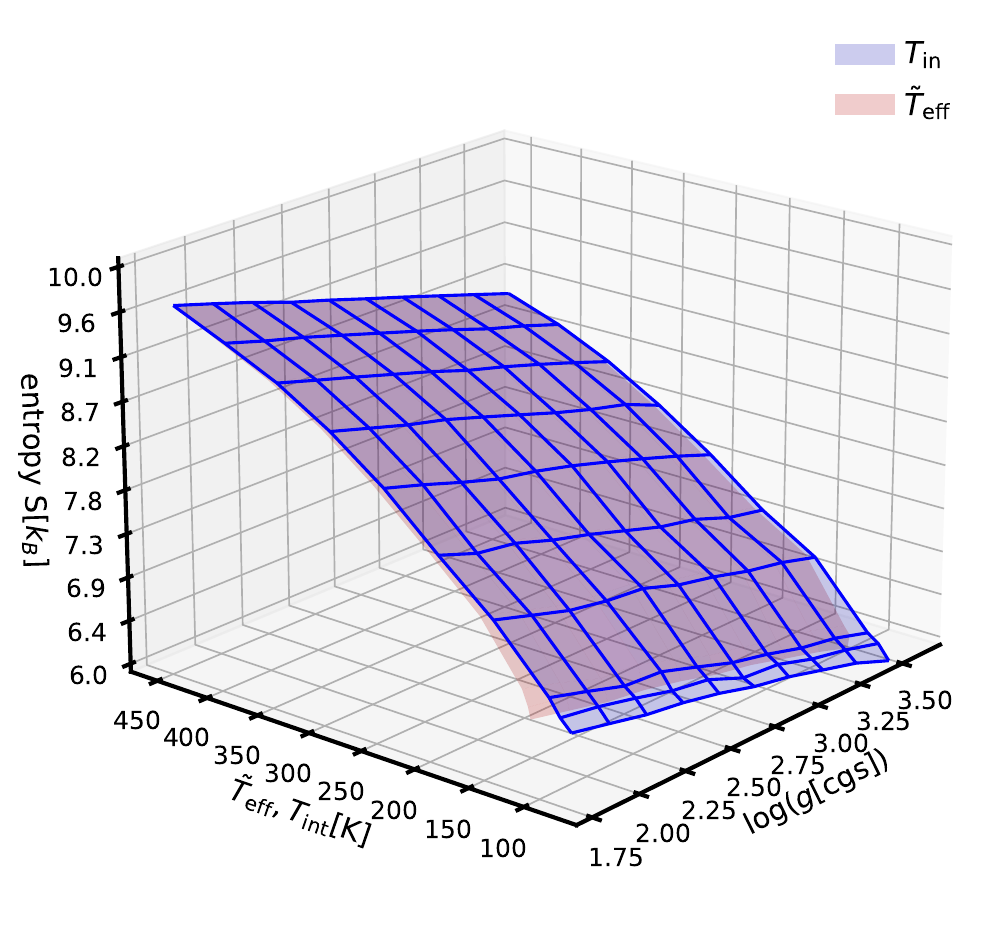}
\caption{The specific entropy in the convective zone under the atmosphere's thin radiative layer, 
plotted as a function of gravity, and effective temperature $\tilde{T}_{\rm eff}$ (red) or internal temperature $T_{\rm int}$ (blue) for our fiducial model ($f=0.5, P_{\rm irr} = 0.15, a_0 = 1\mu$m). At low temperatures, 
$\tilde{T}_{\rm eff}$ starts to become significantly larger than $T_{\rm int}$ due to the increasing contribution of stellar irradiation.}
 \label{fig:surface}
\end{figure*}

To demonstrate its structure, in Figure \ref{fig:surface} we plot the atmospheric entropy per baryon at depth as a function of either $T_{\rm int}$ or $\tilde{T}_{\rm eff}$ and surface gravity using our fiducial boundary table. 
The entropy surface in the $T_{\rm int}, g$ plane is shown in blue (with wireframe) and the entropy surface in the $\tilde{T}_{\rm eff}, g$ plane is shown in red. 
With a planetary evolution code one follows a gas giant along entropy surfaces from high entropy and low $g$ (upper left) to low entropy and high $g$ (lower right), interpolating the table within grid points. 
It's apparent that at high entropy and temperature the surfaces converge, 
while at low temperature $\tilde{T}_{\rm eff}$ starts to deviate from $T_{\rm int}$ due to the increasing contribution of irradiation. 
In the remainder of this paper, we will omit the tilde symbol in $\tilde{T}_{\rm eff}$ for simplicity.

\section{Homogeneous Adiabatic Evolutionary Calculations}
\label{sec:evo_variables}

In traditional adiabatic cooling calculations, 
the internal flux and entropy are related by energy conservation:

\begin{equation}
    \frac{dL}{dm} = -T \frac{dS}{dt},
\end{equation}

where $L$ is the luminosity, $S$ is the specific entropy per mass, and $t$ is the time. 
At a given time, $S$ would be the same throughout an adiabatic envelope and independent of radius. 
$m$ is the integrated mass ($\int^r 4\pi r'^2 \rho(r') dr'$) and $T$ is the local temperature, both functions of shell radius $r$. 
From this equation we can relate the timestep between two models $\Delta t$ that differ in entropy by $\Delta S$. 
The luminosity ($L$) is derived physically from the planet interior and is equal to $4\pi R^2 \sigma T^4_{\rm int}$, 
where the value of $T_{\rm int}$ at a given $S$ and surface gravity is inverted from a model atmosphere table (see Figure \ref{fig:surface}). 
$T_{\rm eff}$ at a given time is also inverted from the model atmosphere table, 
but does not go into the calculation of the variation of internal entropy and the evolution of the temperature profile with time. 


\begin{figure*}[htbp!]
\centering
\includegraphics[width=0.9\textwidth,clip=true]{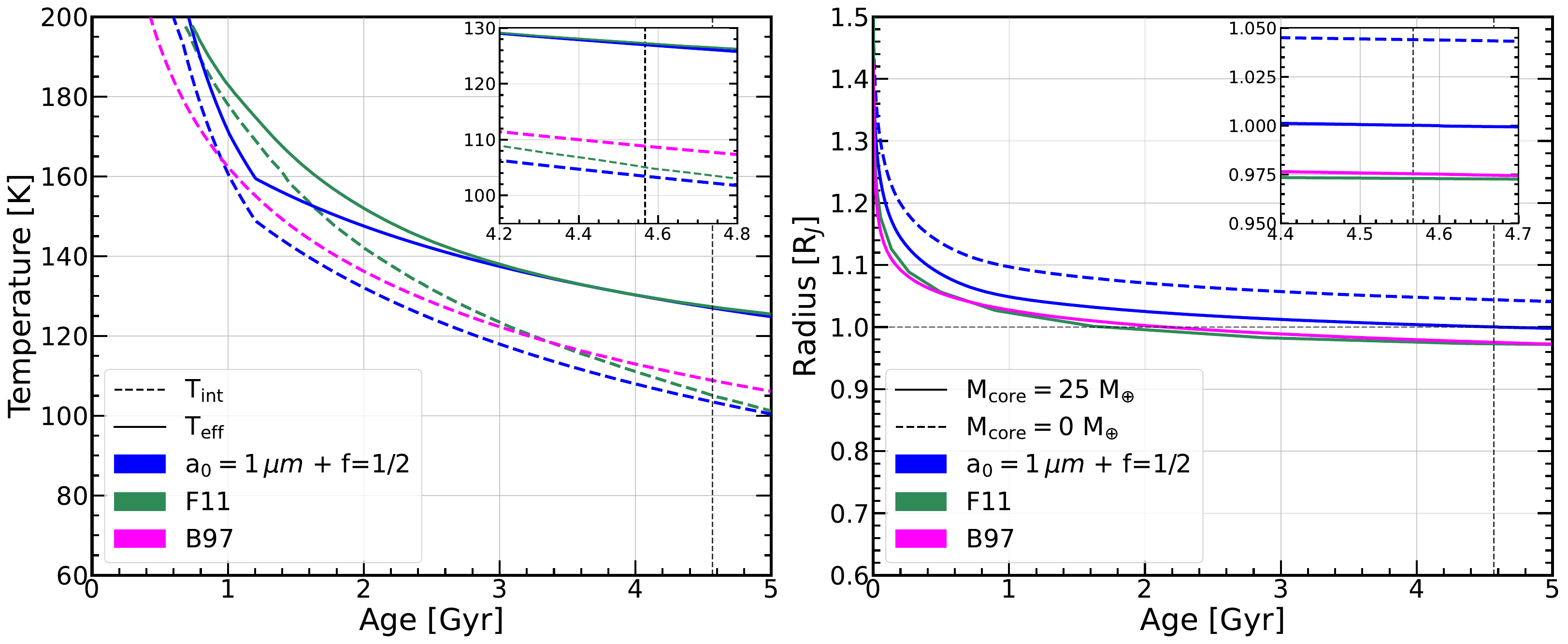}
\caption{Left panel: Evolution of internal brightness temperature $T_{\rm int}$ and emission temperature $T_{\rm eff}$ for our fiducial Jupiter model (blue) evolved in our adiabatic code, 
the \citet{Fortney2011} boundary condition (taken from their Figure 7), and the \citet{Burrows1997} boundary condition evolved in our adiabatic code. 
Right panel: the evolution of the model Jupiter radius. An extra evolutionary track for zero core mass is plotted with a dashed line to show that a solid core is needed to make the planet more compact.}
 \label{fig:core_mass_comp}
\end{figure*}

In Figure \ref{fig:core_mass_comp}, we plot the evolutionary tracks of a ``hot-start" Jupiter for our fiducial boundary conditions, as well as those from the \citet{Burrows1997} and \citet{Fortney2011} models. All models used the \citet{Saumon1995} equation of state. The left panel shows $T_{\rm int}, \tilde{T}_{\rm eff}$ and the right panel shows the radius evolution. The \citet{Fortney2011} evolution track for Jupiter is taken from their Figure 7, 
for which they included a 10$M_\odot$ metal core and approximately 15$M_\odot$ of extra metallicity in the form of water in the H/He adiabatic envelope. 
In their paper, they also used their adiabatic evolution code to calculate the evolution of Jupiter with \citet{Burrows1997} tables that are for $Z = Z_\odot$ and do not include irradiation. 
They obtain a low $T_{\rm int}$ of $\approx 100$K at 4.56 Gyrs (see their Figure 7) \footnote{\citet{Fortney2011} created effective temperatures for the \citet{Burrows1997} boundary condition table through post-processing of irradiation, assuming a constant Bond albedo of 0.343. 
To avoid confusion, we do not adopt this tactic. 
Hence, no corresponding $\tilde{T}_{\rm eff}$s are plotted for the \citet{Burrows1997} calculation.}. 
However, 
we are unable to reproduce this result from \citet{Fortney2011} using the \citet{Burrows1997} boundary condition table at our disposal for any equivalent core/heavy-element mass between $0$ to $30M_\oplus$.  
Instead, 
in Figure \ref{fig:core_mass_comp}, we show an evolution track produced by our code with $25M_\oplus$. This model yields at 4.567 Gyrs a $T_{\rm int}$ $110$ K, which is larger than the $T_{\rm int}$ from \citet{Fortney2011} models, 
but consistent with earlier calculations performed by \texttt{CoolTLusty}. 

Nevertheless, with our fiducial table, and a core mass of 25$M_\oplus$, we obtain $\tilde{T}_{\rm eff} \sim 127K$ at 4.56 Gyrs. 
While $T_{\rm int}$ in our calculation is smaller than found using the \citet{Burrows1997} table, the total emission or effective temperature is reasonably close to the \citet{Lietal2018} estimation. Moreover, adding the fiducial core mass has a pronounced influence on the radius evolution. 
In the right panel of Figure \ref{fig:core_mass_comp}, 
we also compare the evolution of planetary radius with (solid line) and without (dashed line) the core mass for the fiducial boundary condition table. 
Generically, with a metal core or equivalent heavy-element mass in the envelope, 
the gas giant has a larger mean density and is more compact. 
Hereafter, we present all results with a solid core of 25 $M_\oplus$.

\subsection{Dependence on the f factor}

\begin{figure*}[htbp!]
\centering
\includegraphics[width=0.9\textwidth,clip=true]{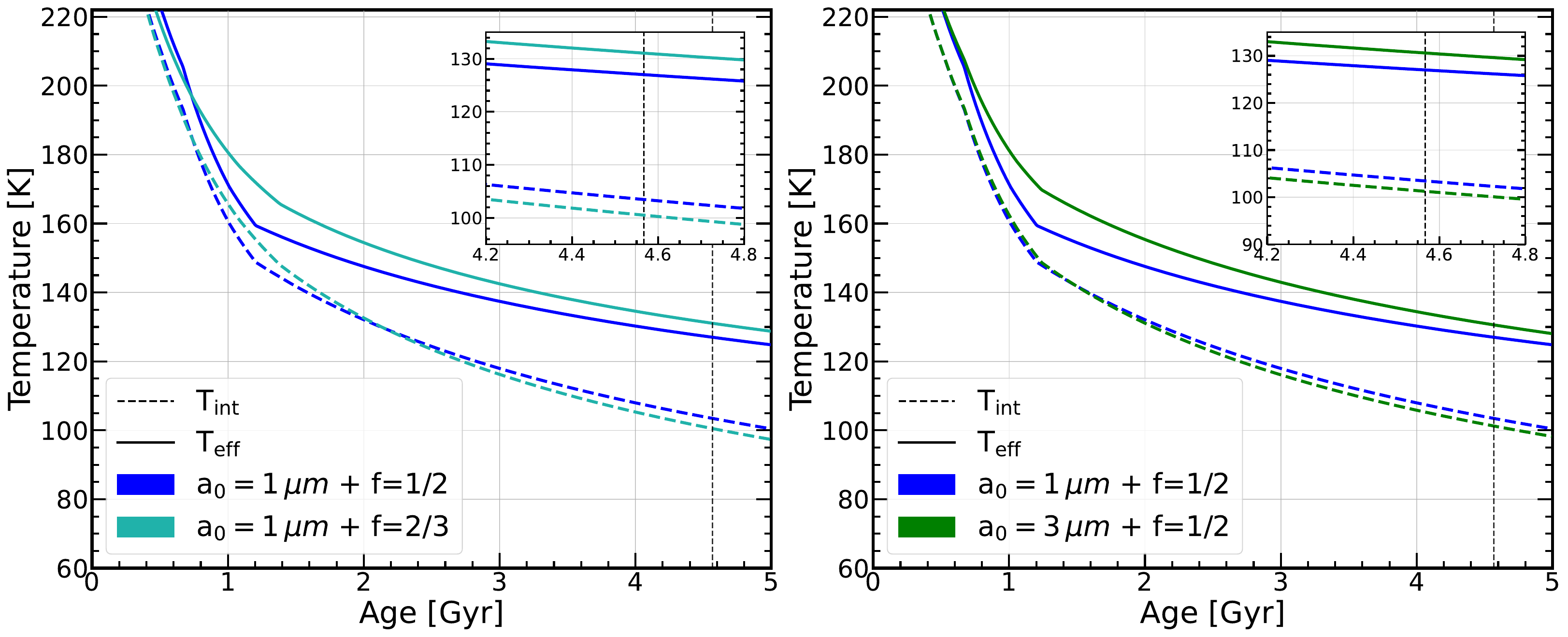}
\caption{Left panel: Jupiter's temperature evolution comparing different zenith angle factors $f$; 
Right panel: Jupiter's temperature evolution comparing different cloud particle size parameters $a_0$.}
 \label{fig:f_comp}
\end{figure*}

In the fiducial models, $f=1/2$ is adopted to approximate the average zenith angle for a moderately irradiated planet. 
For strong irradiation, this factor approaches $2/3$ \citep{Burrows2008} so there is a degree of freedom here not constrained by planar models. 
In the left panel of Figure \ref{fig:f_comp}, we compare $T_{\rm int} $ and $ T_{\rm eff}$ evolutionary tracks using $f=0.67$ and our fiducial tables (keeping $a_0 = 1\mu$m). 
We find that the $T_{\rm int}$ and $T_{\rm eff}$ values at the current age of Jupiter increase and decrease, respectively, with decreasing $f$, 
such that low $f$ evolutionary curves are generally anchored between those of high $f$, although there are some sharp non-linear transitions at $T_{\rm int} \sim $150-200K due to the onset of ammonia clouds.
This is consistent with the expectation that these temperatures should converge towards the isolated planet case with $T_{\rm eff} = T_{\rm int}$ (similar to the \citet{Burrows1997} evolutionary curve) when $f$ approaches zero (i.e., when irradiation is ignored). Interestingly, 
a decrease in $f$ may also generally represent the effect of an extra scatterer (\S\ref{sec:albedo}). 
If the fact that $A_B > A_G$ is a consequence of $q>1$, then one can decrease $f$ by a factor of $\approx (A_B - A_G)/(1-A_G)$ to account for this reduction. 
This will result in a larger $T_{\rm int}$ at 4.56 Gyrs, in the direction of being slightly more consistent with \citet{Lietal2018}. 
However, the inclusion of a ``phase-integral correction" in $f$ is quite artificial, 
and we certainly do not expect $q$ or this modification of the scattering fraction to be constant in time,
Nevertheless, 
this crude $f$ pseudo-dependence illuminates to zeroth-order the dependence of $T_{\rm eff}$ and $T_{\rm int}$ on the phase integral, 
such that for multi-dimensional radiative transfer codes that do include this effect, 
we expect evolutionary tracks to move slightly closer to the observations.

\subsection{Dependence on Particle Size}

In the right panel of Figure \ref{fig:f_comp}, we display the evolution of $T_{\rm int}$ and $T_{\rm eff}$ for fiducial and $a_0 = 3 \mu$m models (fixing $f=1/2$). 
Just as with the dependence on smaller $f$, for smaller particle size, 
$T_{\rm int}$ at 4.56 Gyrs is larger, 
but $T_{\rm eff}$ is not necessarily also larger; 
this results from the smaller contribution from irradiation due to larger albedo. 
For larger particle sizes, the two temperatures tend to diverge, 
and the contrast between $T_{\rm int}$ and $T_{\rm eff}$ becomes large. 
In the small-particle-size limit, 
for a negligible irradiation contribution (when incident radiation is completely scattered), 
we expect these temperatures to converge, similar to what is found in the $f\rightarrow 0$ limit.

\subsection{Dependence on Abundances}

All of the models above are for $Y = 0.25$ and $Z = 3.16Z_\odot$ in the atmosphere.  We also compare evolutionary tracks with $Y = 0.25, Z = 10Z_\odot$ and $Y = 0.22, Z = 3.16Z_\odot$
and find that varying $Y$ in the atmosphere has little or no effect on the planetary evolution. 
However, an increase in $Z$ leads to an increase in $T_{\rm int}$ and $T_{\rm eff}$ and slows planetary cooling by raising the atmospheric opacity. 
This effect arises even at earlier times, unlike the effect of ammonia clouds which appears only at late stages. 
Note again that these abundance variations are in the atmosphere, 
and that our internal adiabats still have $Y = 0.27, Z = 3.16Z_\odot$ fixed. 
The dependence of gas-giant evolution on internal compositions and their profiles (e.g. effect of helium depletion) is properly a subject of future work.

\section{Conclusion} \label{sec:discussions}

In this study, we develop a set of atmospheric models using the 1D atmosphere and spectral code \texttt{CoolTLusty}. 
Our goal is to create state-of-the-art boundary condition tables that could be used for studying the evolution of gas giant planets. 
The atmospheric opacities employed were significantly updated by \citet{Lacy2023} and we implement realistic treatments of clouds and irradiation to calibrate our parameters with the observed temperature structure and albedo spectrum of Jupiter. 
We simulated the internal evolution of a ``hot-start" Jupiter using this set of tables in the context  of the traditional adiabatic paradigm. 
This approach is being challenged by the new {Juno} data \citep{Wahl2017,Bolton2017}, 
but the viability of these atmospheric boundary conditions for any evolutionary model is not compromised. 
We find that with reasonable irradiation and cloud parameters we obtain an atmospheric boundary condition table that cools down a ``hot-start" Jupiter to close to its current measured thermal state with its measured geometric albedos. 

In addition to providing a useful atmosphere reference model for
future giant-planet cooling calculations, we explored the dependence on various physical parameters. 
The average zenith angle parameter $f$ and cloud modal particle size $a_0$ have  particular sway over the influence of irradiation. 
If the absorption fraction of irradiation is very high, 
then the effective temperature $T_{\rm eff}$ may not be able to cool down to near the measured value of $\sim$125 K at Jupiter's current age. 
However, by using fiducial sets of cloud parameters to ensure a reasonable geometric albedo spectrum, 
we observed that $T_{\rm eff}$ cools down faster than observed using previous approaches \citep{Burrows1997,Fortney2011}, 
better matching the surface observations of Jupiter. 
The average zenith angle also has an impact, and for a slight reduction in $f$, $T_{\rm eff}$ also cools down a bit faster. 
In addition, the atmospheric metallicity affects the cooling rate, 
with a higher $Z$ raising the atmosphere opacity and resulting in slower cooling. 
The helium fraction $Y$ in the atmosphere has a minimal effect. 
For this study, we fixed the stellar luminosity and implemented only ammonia clouds. 
Nevertheless, 
we believe incorporating time-changing stellar insolation and allowing for the formation of water clouds at higher atmospheric temperatures should have a less significant effect than ammonia clouds, 
since a giant's initial evolution across this parameter space is much more rapid than late-stage cooling.

Our collection of tables, calibrated using data from Jupiter 
and covering a relatively broad range of parameters that includes varying $Z$, $a_0$, and $f$, serves as a foundational resource for advancing our understanding of the evolution of Jupiter-like exoplanets and for modeling their spectral evolution. 
This new boundary-condition dataset is available to the scientific community for future research into the evolution of both solar-system giants and giant exoplanets. 
Nevertheless, the effects of the composition gradients inferred from {Juno} data \citep{Wahl2017,Bolton2017} and of helium rain \citep{Stevenson1975, Stevenson1977, Fortney2004, Mankovich2016, Mankovich2020}, 
as well as updates to the H/He equation of state \citep{Nettelmann2012,MH2013,Miguel2016,Howard2023a,Howard2023b}, 
demand a more generalized view beyond the adiabatic paradigm. 
We used the latter here merely to test our atmosphere calculations in light of previous work. 
In summary, these new boundary tables and atmospheres are meant to support the next generation of comprehensive giant planet models, which is already well underway \citep{Nettelmann2015,Pustow2016,Vazan2016,Mankovich2016,Vazan2018,Mankovich2020,Vazan2023}.



\begin{acknowledgments}
We thank Brianna Lacy for updated opacity tables and helpful discussions. 
Funding (or partial funding) for this research was provided by the Center 
for Matter at Atomic Pressures (CMAP), a National Science Foundation (NSF) 
Physics Frontier Center, under Award PHY-2020249. Any opinions, findings, 
conclusions or recommendations expressed in this material are those of the 
author(s) and do not necessarily reflect those of the National Science 
Foundation.
\end{acknowledgments}

%

\vspace{5mm}


\software{\texttt{CoolTLusty} \citep{Hubeny1995,Sudarsky2000,Sudarsky2003,Sudarsky2005,Burrows2008}. Relevant boundary condition entropy tables used for generating our figures are presented at https://doi.org/10.5281/zenodo.8297690.
}



\appendix


\bibliography{sample631}{}
\bibliographystyle{aasjournal}



\end{document}